\documentclass[twocolumn,floatfix,aps,superscriptaddress]{revtex4}
\usepackage{graphicx}
\usepackage{amsmath}
\usepackage{amssymb}
\usepackage{bm}
\usepackage{hyperref}
\usepackage{mathtools} 

\usepackage{color}

\begin{document}
\newcommand{\overbrack}{\overbracket[0.5pt][1.5pt]} 

\title{Localization landscape for Dirac fermions}
\author{G. Lemut}
\affiliation{Instituut-Lorentz, Universiteit Leiden, P.O. Box 9506, 2300 RA Leiden, The Netherlands}
\author{M. J. Pacholski}
\affiliation{Instituut-Lorentz, Universiteit Leiden, P.O. Box 9506, 2300 RA Leiden, The Netherlands}
\author{O. Ovdat}
\affiliation{Instituut-Lorentz, Universiteit Leiden, P.O. Box 9506, 2300 RA Leiden, The Netherlands}
\author{A. Grabsch}
\affiliation{Instituut-Lorentz, Universiteit Leiden, P.O. Box 9506, 2300 RA Leiden, The Netherlands}
\author{J. Tworzyd{\l}o}
\affiliation{Institute of Theoretical Physics, Warsaw University, Ho\.{z}a 69, 00--681 Warsaw, Poland}
\author{C. W. J. Beenakker}
\affiliation{Instituut-Lorentz, Universiteit Leiden, P.O. Box 9506, 2300 RA Leiden, The Netherlands}
\date{January 2020}
\begin{abstract}
In the theory of Anderson localization, a landscape function predicts where wave functions localize in a disordered medium, without requiring the solution of an eigenvalue problem. It is known how to construct the localization landscape for the scalar wave equation in a random potential, or equivalently for the Schr\"{o}dinger equation of spinless electrons. Here we generalize the concept to the Dirac equation, which includes the effects of spin-orbit coupling and allows to study quantum localization in graphene or in topological insulators and superconductors. The landscape function $u(\bm{r})$ is defined on a lattice as a solution of the differential equation $\overbrack{{H}}u(\bm{r})=1$, where $\overbrack{{H}}$ is the Ostrowsky comparison matrix of the Dirac Hamiltonian. Random Hamiltonians with the same (positive definite) comparison matrix have localized states at the same positions, defining an equivalence class for Anderson localization. This provides for a mapping between the Hermitian and non-Hermitian Anderson model.   
\end{abstract}
\maketitle

\emph{Introduction ---} The localization landscape is a new tool in the study of Anderson localization, pioneered in 2012 by Filoche and Mayboroda \cite{Fil12}, which has since stimulated much computational and conceptual progress \cite{Fil13,Arn16,Ste17,Fil17,Pic17,Li17,Cha19,Arn19,Har18,Bee19}. The ``landscape'' of a Hamiltonian $H$ is a function $u(\bm{r})$ that provides an upper bound for eigenstates $\psi$ at energy $E>0$:
\begin{equation}
 |\psi(\bm{r})|/|\psi|_{\rm max}\leq E\,u(\bm{r}),\;\;|\psi|_{\rm max}=\textstyle{\max_{\bm{r}}}|\psi(\bm{r})|.\label{landscapeineq}
\end{equation}
This inequality implies that a localized state is confined to spatial regions where $u\gtrsim 1/E$. Extensive numerical simulations \cite{Arn19} confirm the expectation that higher and higher peaks in $u$ identify the location of states at smaller and smaller $E$.

Such a predictive power would be unremarkable for particles confined to potential wells (deeper and deeper wells trap particles at lower and lower energies). But Anderson localization happens because of wave interference in a random ``white noise'' potential, and inspection of the potential landscape $V(\bm{r})$ gives no information on the localization landscape $u(\bm{r})$.

Filoche and Mayboroda considered the localization of scalar waves, or equivalently of spinless electrons, governed by the Schr\"{o}dinger Hamiltonian $H=-\nabla^2+V$. They used the maximum principle for elliptic partial differential equations to derive \cite{Fil12} that the inequality \eqref{landscapeineq} holds if $V>0$ and $u$ is the solution of
\begin{equation}
[-\nabla^2+V(\bm{r})]u(\bm{r})=1.\label{u_FM}
\end{equation}
Our objective here is to generalize this to spinful electrons, to include the effects of spin-orbit coupling and study localization of Dirac fermions.

\emph{Construction of the landscape function ---}
Our key innovation is to use Ostrowski's comparison matrix \cite{Ost37,Ost56,nomenclature,Ber94} as a general framework for the construction of a localization landscape on a lattice. By definition, the comparison matrix $\overbrack{{H}}$ of a complex matrix ${{H}}$ has elements 
\begin{equation}
{\overbrack{{H}}}_{nm}=\begin{cases}
|{{H}}_{nn}|&{\rm if}\;\;n=m,\\
-|{{H}}_{nm}|&{\rm if}\;\;n\neq m.
\end{cases}\label{compmatrixdef}
\end{equation}
In our context the index $n=1,2,\ldots$ labels both the discrete space coordinates as well as any internal (spinor) degrees of freedom. The comparison theorem \cite{Ost37} states that if the comparison matrix is positive-definite, then \cite{note1}
\begin{equation}
|{{H}}^{-1}|\leq{\overbrack{{H}}}\,^{-1},\label{Hcomparison}
\end{equation}
where both the absolute value and the inequality is taken elementwise. 

We apply Eq.\ \eqref{Hcomparison} to an eigenstate $\Psi$ of ${{H}}$ at energy $E$,
\begin{align}
|E^{-1}\Psi_n|&=|({{H}}^{-1}\Psi)_n|\leq\textstyle{\sum_{m}}\bigl| \bigl({{H}}\,^{-1}\bigr)_{nm}\bigr||\Psi_m|\nonumber\\
&\leq|\Psi|_{\rm max}\textstyle{\sum_{m}}\bigl({\overbrack{{H}}}\,^{-1}\bigr)_{nm},\label{comparisonineq1}
\end{align}
with $|\Psi|_{\rm max}=\max_n|\Psi_n|$. We now define a landscape function $u$ with elements $u_n$ in terms of a set of linear equations with coefficients given by the comparison matrix:
\begin{equation}
\overbrack{H}u=1\Leftrightarrow\textstyle{\sum_{m}}{\overbrack{{H}}}_{nm}u_m=1,\;\;n=1,2,\ldots N,\label{u_comparison}
\end{equation}
which implies that
\begin{equation}
\textstyle{\sum_{m}}\bigl({\overbrack{{H}}}\,^{-1}\bigr)_{nm}= u_n.
\end{equation}
Substitution into Eq.\ \eqref{comparisonineq1} thus gives the desired inequality
\begin{equation}
|\Psi_n|/|\Psi|_{\rm max}\leq  |E|\, u_n.\label{landscapeineqdiscrete}
\end{equation}

As a sanity check, we make contact with the original landscape function \cite{Fil12} for the Schr\"{o}dinger Hamiltonian $H_{\rm S}=p^2/2m+V$, with $V>0$. The Laplacian is discretized in terms of nearest-neighbor hoppings on a lattice. For each dimension
\begin{equation}
\begin{split}
&p^2\mapsto (\hbar/a)^2(2-2\cos ka)\Rightarrow\\
&(H_{\rm S})_{nm}=t_0(2\delta_{nm}-\delta_{n-1,m}-\delta_{n+1,m})+V_n\delta_{nm},
\end{split}
\label{p2discretization}
\end{equation}
with lattice constant $a$ and hopping matrix element $t_0=\hbar^2/2ma^2$. The comparison matrix ${\overbrack{H}}_{\rm S}$ is equal to $H_{\rm S}$ and is positive-definite, so that Eq.\ \eqref{u_comparison} is a discretized version of the original landscape equation $H_{\rm S}u=1$ \cite{Fil12,Lyr15}.

\emph{Rashba Hamiltonian ---}
Our first novel application is to introduce spin-orbit coupling of the Rashba form,
\begin{equation}
H_{\rm R}=H_{\rm S}+\tfrac{1}{2}\{\lambda, p_x\}\sigma_y-\tfrac{1}{2}\{\lambda,p_y\}\sigma_x.\label{HRdef}
\end{equation}
(The anticommutator $\{\cdots\}$ enforces Hermiticity when $\lambda$ is spatially dependent.) The comparison matrix is now no longer equal to the Hamiltonian, in 1D one has
\begin{equation}
({\overbrack{H}}_{\rm R})_{ij}=(H_{\rm S})_{ij}-\frac{\hbar}{4a}|\lambda_i+\lambda_j|(\delta_{i-1,j}+\delta_{i+1,j})\sigma_x.\label{HR1D}
\end{equation}
The $i,j,$ indices label the spatial positions, the spinor indices are implicit in the Pauli matrix.

\begin{figure}[tb]
\centerline{\includegraphics[width=0.8\linewidth]{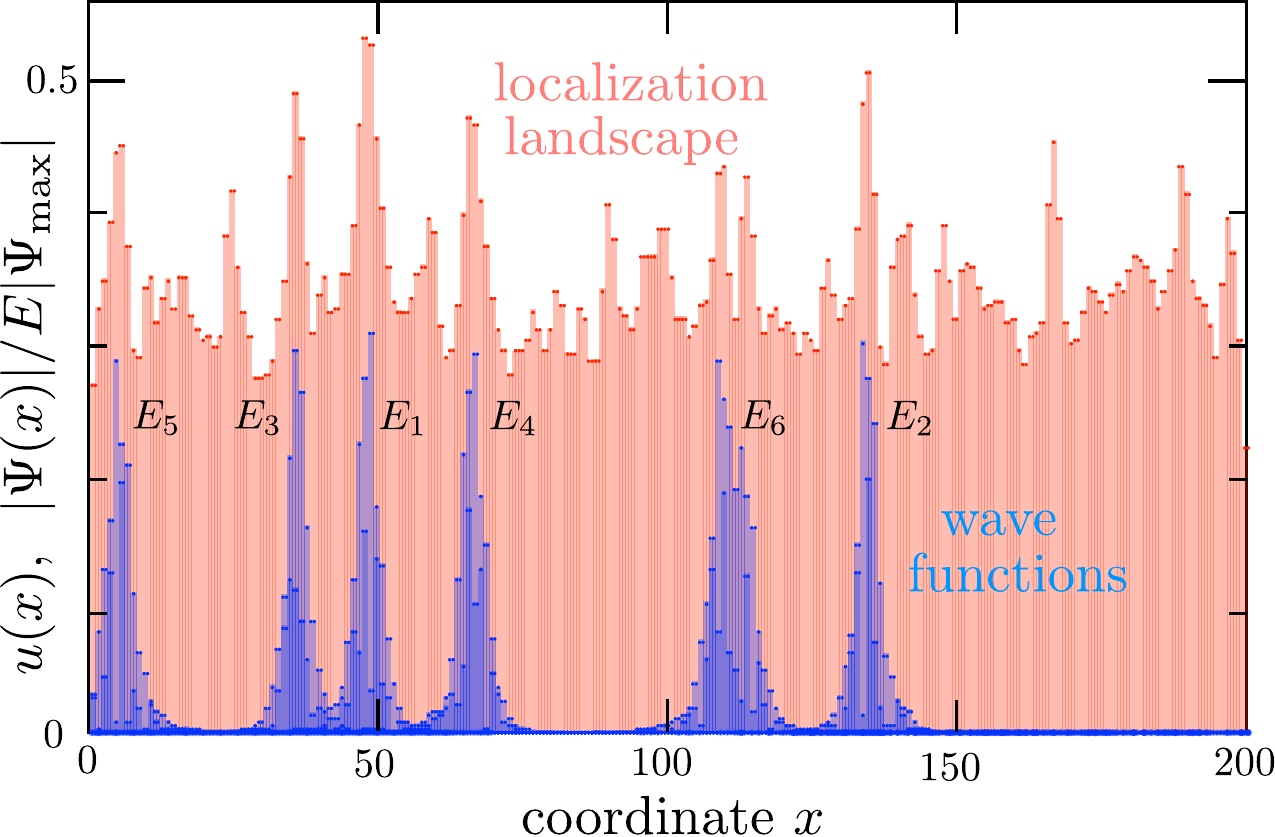}}
\caption{Landscape function $u(x)$ (red) and normalized wave function profile $|\Psi(x)|/E|\Psi_{\rm max}|$ (blue) for the 6 lowest (twofold degenerate) eigenstates of the disordered 1D Rashba Hamiltonian \eqref{HR1D} (parameters $V_0=4t_0$, $\bar\lambda=0$, $\delta\lambda=3\hbar/a$, hard-wall boundary conditions). The 1D array has $n=1,2,\ldots 200$ sites, in the plot $x=n$ shows the first spinor component and $x=n+1/2$ shows the second spinor component. The wave functions are labeled by the corresponding energy levels $\{E_1,\ldots E_6\}=\{3.273, 3.3371, 3.414, 3.446, 3.508, 3.516\}$ (in units of $t_0$).}
\label{fig_Rashba1D}
\end{figure}

\begin{figure}[tb]
\centerline{\includegraphics[width=1\linewidth]{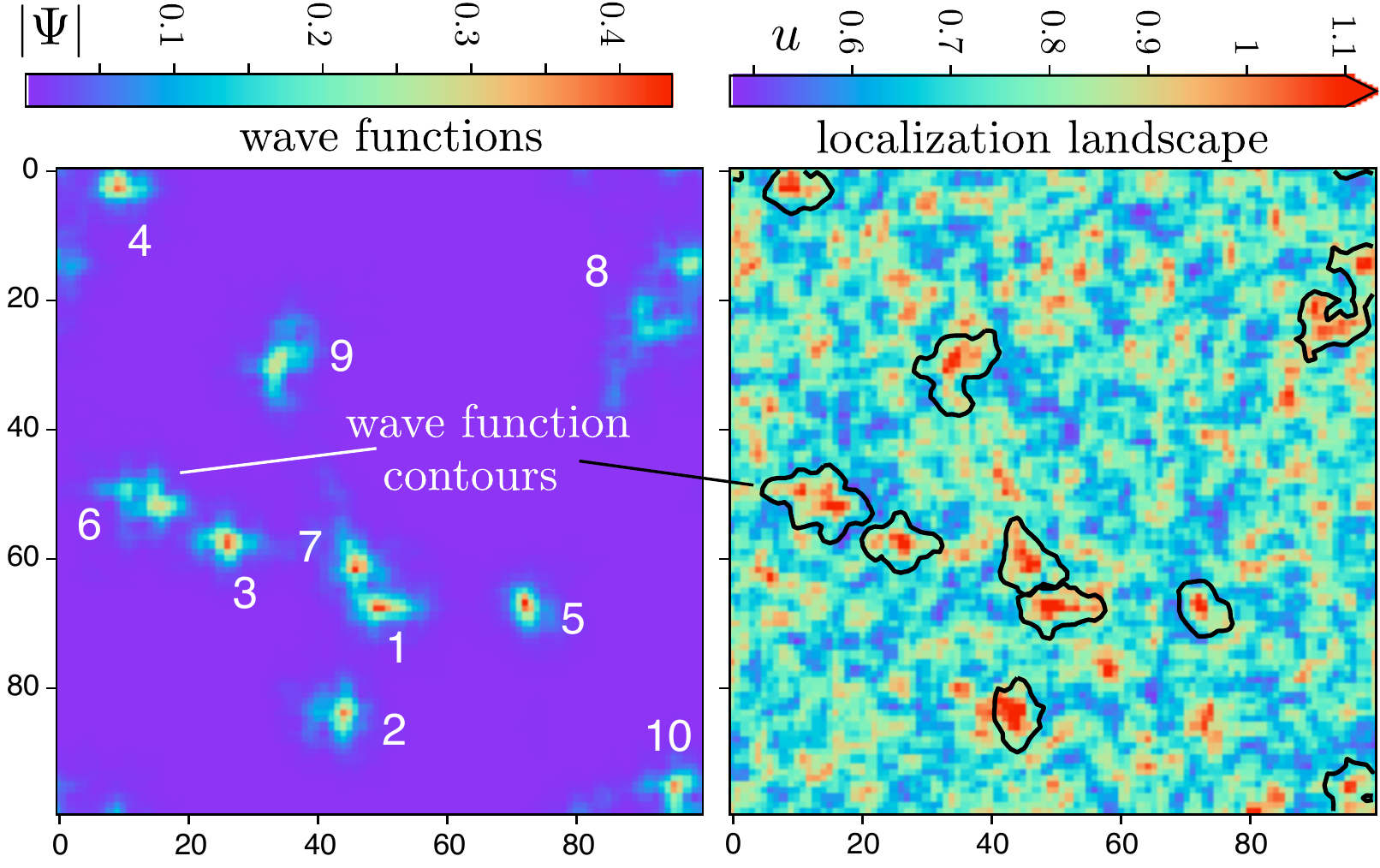}}
\caption{Same comparison as in Fig.\ \ref{fig_Rashba1D}, but now for the 2D Rashba Hamiltonian, discretized on a $100\times 100$ square lattice (parameters $V_0=6t_0$, $\bar\lambda=2\delta\lambda=2\hbar/a$, periodic boundary conditions).The left panel shows the spinor norm $|\Psi_n(\bm{r)}|$ for the 10 lowest (twofold degenerate) eigenstates of ${{H}}_{\rm R}$. The right panel shows the localization landscape. The black contours (computed at 10\% of the peak height of $|\Psi|$) identify the location of the 10 eigenstates --- to show the close correspondence with the local maxima of $u(\bm{r})$.
}
\label{fig_Rashba2D}
\end{figure}

As a test, to isolate the effect of spin-orbit coupling, we place all the disorder in the Rashba strength $\lambda_n$, which fluctuates randomly from site to site, uniformly in the interval $(\bar{\lambda}-\delta\lambda,\bar{\lambda}+\delta\lambda)$. The electrostatic potential is a constant offset $V_0$, chosen sufficiently large that ${\overbrack{H}}_{\rm R}$ is positive-definite \cite{note2}. Examples in 1D and in 2D are shown in Figs.\ \ref{fig_Rashba1D} and \ref{fig_Rashba2D}. The highest peaks in the landscape function match well with the lowest eigenfunctions. 

\emph{Dirac Hamiltonian ---}
We next turn to Dirac fermions, first in 1D. The Dirac Hamiltonian
\begin{equation}
H_{\rm D}=v_{\rm F}p_x\sigma_x+V\sigma_0+\mu\sigma_z\label{HD1D}
\end{equation}
contains a scalar potential $V$ proportional to the $2\times 2$ unit matrix $\sigma_0$ and a staggered potential $\mu$ proportional to $\sigma_z$, acting on the two-component wave function $\Psi=(\psi_A,\psi_B)$. This would apply to a graphene nanoribbon on a substrate such as hexagonal boron nitride, which differentiates between the two carbon atoms in the unit cell without causing intervalley scattering \cite{Gio07}. 

The symmetric discretization $\partial_x\Psi\mapsto(1/2a)[\Psi(x+a)-\Psi(x-a)]$ suffers from fermion doubling \cite{Sta82,Two08} --- it corresponds to a $\sin ka$ dispersion with a second species of massless Dirac fermions at the edge of the Brillouin zone ($k=\pi/a$). To avoid this, and restrict ourselves to a single valley, we use a staggered-fermion discretization \textit{a la} Susskind \cite{Sus77,Her12}:
\begin{equation}
p_x\sigma_x\Psi\mapsto(-i\hbar/a)\begin{pmatrix}
\psi_{B}(x)-\psi_{B}(x-a)\\
\psi_{A}(x+a)-\psi_{A}(x)
\end{pmatrix}.\label{eq_staggering}
\end{equation}
The corresponding dispersion \cite{note4}
\begin{equation}
E(k)=\pm t_1\sqrt{2-2\cos ka},\;\;t_1=\hbar v_{\rm F}/a,\label{dispersionrelation}
\end{equation}
has massless fermions only at the center of the Brillouin zone ($k=0$).

\begin{figure}[tb]
\centerline{\includegraphics[width=0.9\linewidth]{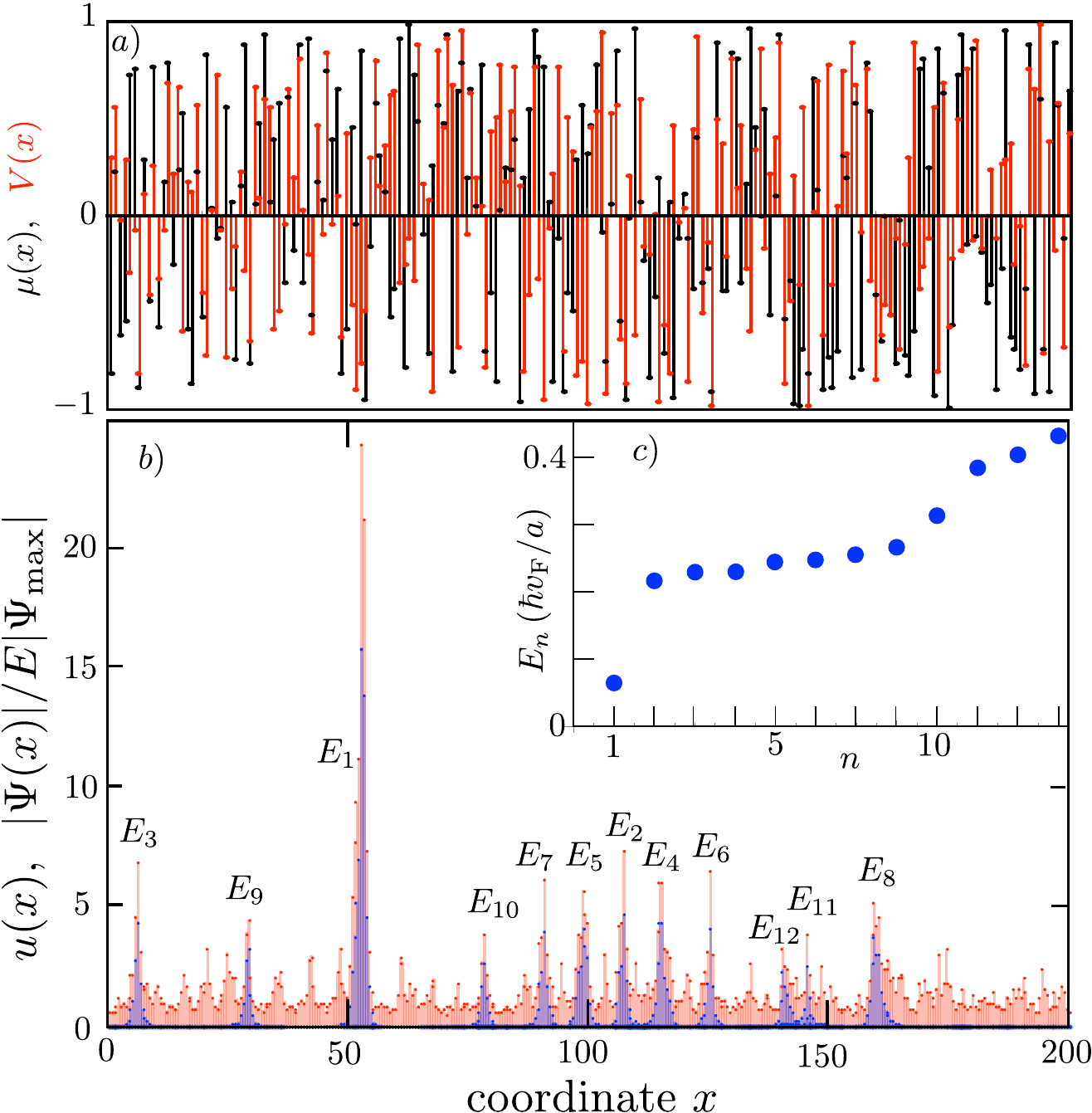}}
\caption{Panel \textit(a): Random scalar potential $V(x)$ (red) and staggered potential $\mu(x)$ (black) for the 1D Dirac Hamiltonian \eqref{HD1D} (parameters $\bar{V}=3t_1$,  $\bar{\mu}=0$, $\delta V=\delta\mu=t_1$, hard-wall boundary conditions). Panel \textit{b)}: Corresponding localization landscape (red) and eigenfunctions of the 12 lowest energy levels (blue), at energies $E_n$ near the band edge plotted in the inset (panel \textit{c}). The peaks in the localization landscape are not correlated in any obvious way with the random potentials, but they accurately predict the location of the low-lying modes.
}
\label{fig_Dirac1D}
\end{figure}

\begin{figure}[tb]
\centerline{\includegraphics[width=0.8\linewidth]{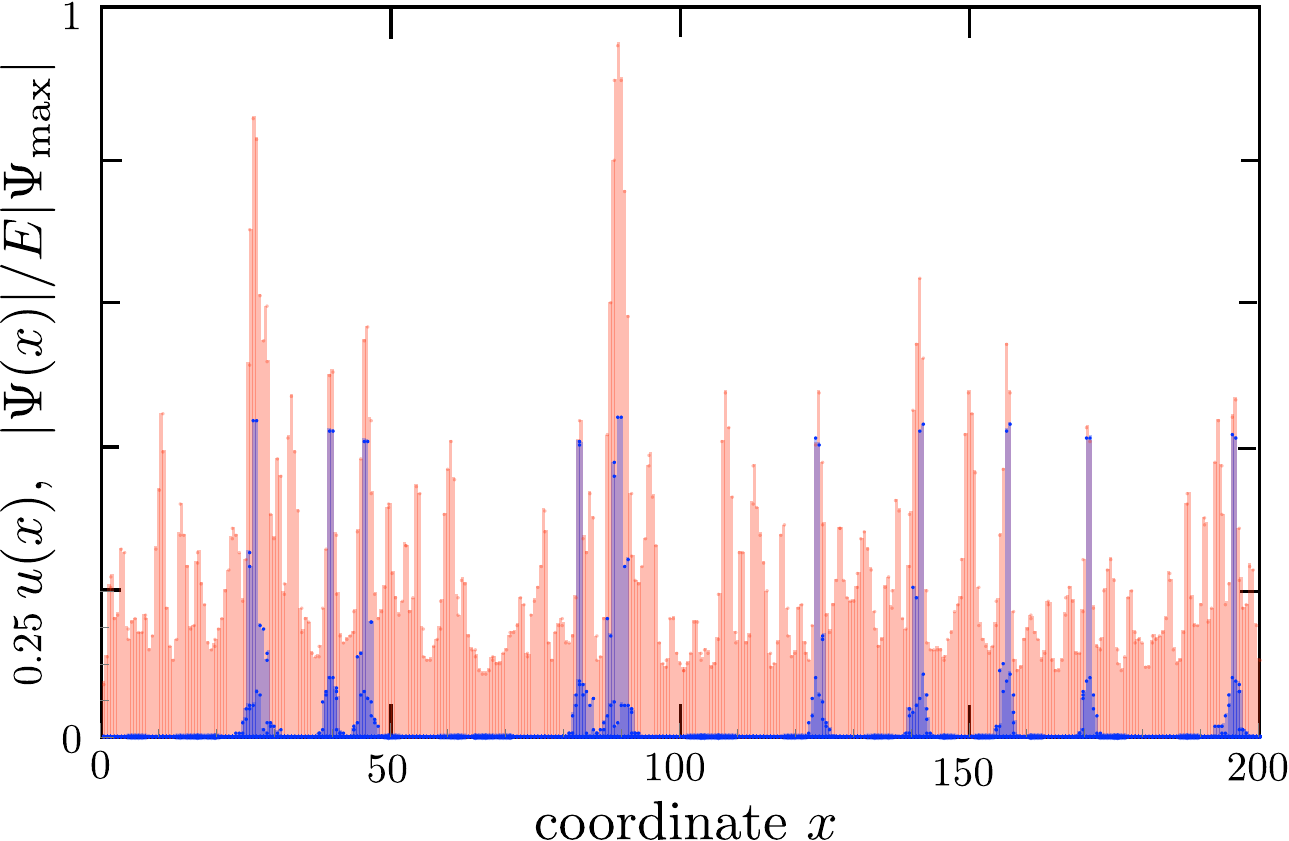}}
\caption{Same as Fig.\ \ref{fig_Dirac1D}b, but now for a gapped system ($\bar{V}=\delta V=0$, $\bar{\mu}=3.5\,t_1$, $\delta\mu=1.5\,t_1$). The eigenfunctions of the 20 levels closest to the gap are shown (blue, $2.3\,t_1<|E_n|<2.5\,t_1$). There are only 10 distinct peaks, because of an approximate $\pm E$ symmetry. The landscape function (red, rescaled by a factor $1/4$) accurately identifies the location of the states near the gap.
}
\label{fig_Dirac1Dextra}
\end{figure}

The comparison matrix takes the form
\begin{equation}
({\overbrack{{H}}}_{\rm D})_{ij}=\begin{pmatrix}
|V_i+\mu_i|\delta_{ij}&-t_1(\delta_{ij}+\delta_{i+1,j})\\
-t_1(\delta_{ij}+\delta_{i-1,j})&|V_i-\mu_i|\delta_{ij}
\end{pmatrix}.
\end{equation}
We take random $V(x)\in(\bar{V}-\delta V,\bar{V}+\delta V)$ and $\mu(x)\in(\bar{\mu}-\delta\mu,\bar{\mu}+\delta\mu)$, chosen independently and uniformly at each lattice site. The condition $|V_i\pm\mu_i|>2t_1$ ensures a positive-definite ${\overbrack{{H}}}_{\rm D}$. As shown in Figs.\ \ref{fig_Dirac1D} and \ref{fig_Dirac1Dextra}, the landscape function computed from ${\overbrack{{H}}}_{\rm D}u=1$ again accurately identifies the locations of the low-lying eigenfunctions (near the band edge in Fig.\ \ref{fig_Dirac1D} and near the gap in Fig.\ \ref{fig_Dirac1Dextra}).

\begin{figure}[tb]
\centerline{\includegraphics[width=1\linewidth]{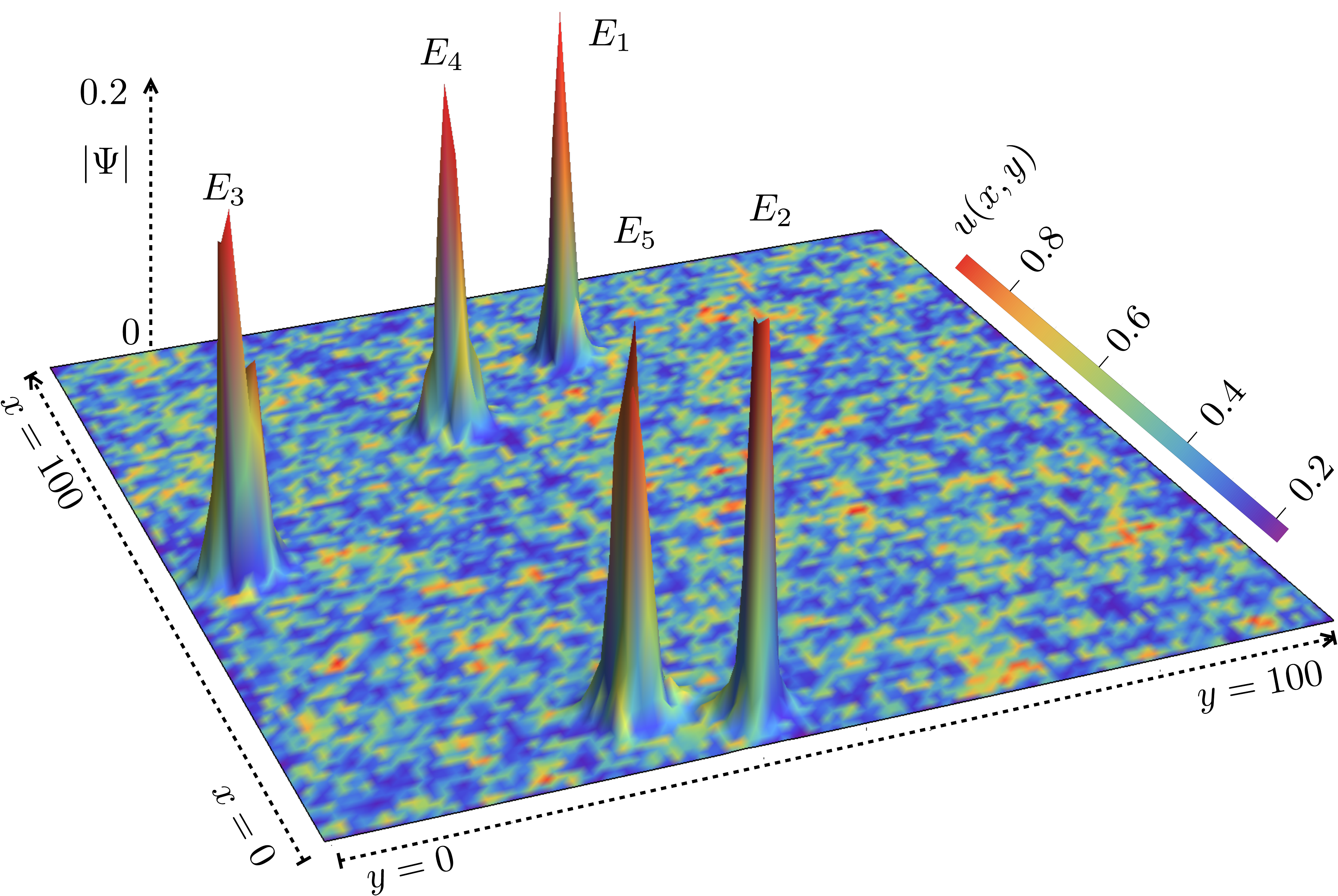}}
\caption{Comparison of the landscape function (2D color scale plot) with wave function amplitudes (3D profile) of the chiral \textit{p}-wave superconductor with Hamiltonian \eqref{HBdGdef} (parameters $\Delta=1$, $\bar{V}=6$, $\delta V=4$, in units of $t_0=\hbar^2/2ma^2$). The wave functions show the five Andreev levels with smallest $E_n>0$ ($E_1,E_2,\ldots E_5=3.763,3.799,3.875,3.882,3.893$). (The charge-conjugate states at $-E_n$ have the same spinor amplitude $|\Psi|$.) The colors of the wave function profile correspond to the landscape function, so a red wave function peak indicates that $u(x,y)$ peaks at the same position.
}
\label{fig_Dirac2D}
\end{figure}

For the 2D Dirac equation we consider a chiral \textit{p}-wave superconductor, with Bogoliubov-De Gennes Hamiltonian \cite{Bee16}
\begin{equation}
{{H}}_{\rm BdG}=\Delta(p_x\sigma_x+p_y\sigma_y)+(V+p^2/2m)\sigma_z.\label{HBdGdef}
\end{equation}
The Pauli matrices act on the electron-hole degree of freedom of a Bogoliubov quasiparticle, and the Hamiltonian is constrained by particle-hole symmetry: $\sigma_x H_{\rm BdG}\sigma_x=-H_{\rm BdG}^\ast$. (A scalar offset $\propto\sigma_0$ is thus forbidden.) The pair potential $\Delta$ opens a gap in the spectrum in the entire Brillouin zone, provided that the electrostatic potential $V$ is nonzero. The gap-closing transition at $V=0$ is a topological phase transition \cite{Rea00}.

We take a uniform real $\Delta$ (no vortices) and a disordered $V(x,y)$, fluctuating randomly from site to site in the interval $(\bar{V}+\delta V,\bar{V}-\delta V)$. Positive $V$ ensures we do not cross the gap-closing transition, so we will not be introducing Majorana zero-modes \cite{Wim10} (the levels are Andreev bound states). Unlike in the case of graphene we can use the symmetric discretization $p\mapsto \sin ka$ --- there is no need for a staggered discretization because the kinetic energy $p^2\mapsto 2-2\cos ka$ prevents fermion doubling at $k=\pi/a$. Results are shown in Fig.\ \ref{fig_Dirac2D}.

\emph{Equivalence classes ---} In the final part of this paper we move beyond applications to address a conceptual implication of the theory. Two complex matrices $A,B$ are called equimodular if $|A_{nm}|=|B_{nm}|$. By the construction \eqref{compmatrixdef}, they have the same comparison matrix, $\overbrack{A}=\overbrack{B}$, and therefore the same landscape function $u_A=u_B$, uniquely determined by the same equation $\overbrack{A}u_A=1=\overbrack{B}u_B$. We thus obtain an equivalence class for Anderson localization: \textit{Equimodular Hamiltonians have localized states at the same position, identified by peaks in the landscape function.}

We have checked this for the 2D Rashba Hamiltonian \eqref{HRdef}: Randomly varying the sign of the coefficient $\lambda(\bm{r})$ from site to site shifts the energy levels around, but the states remain localized at the same positions. More generally, one could try to vary the coefficients over the complex plane, preserving the norm. This would produce a non-Hermitian eigenvalue problem, and one might wonder whether the whole approach breaks down. It does not, as we will now demonstrate.

\begin{figure}[tb]
\centerline{\includegraphics[width=0.9\linewidth]{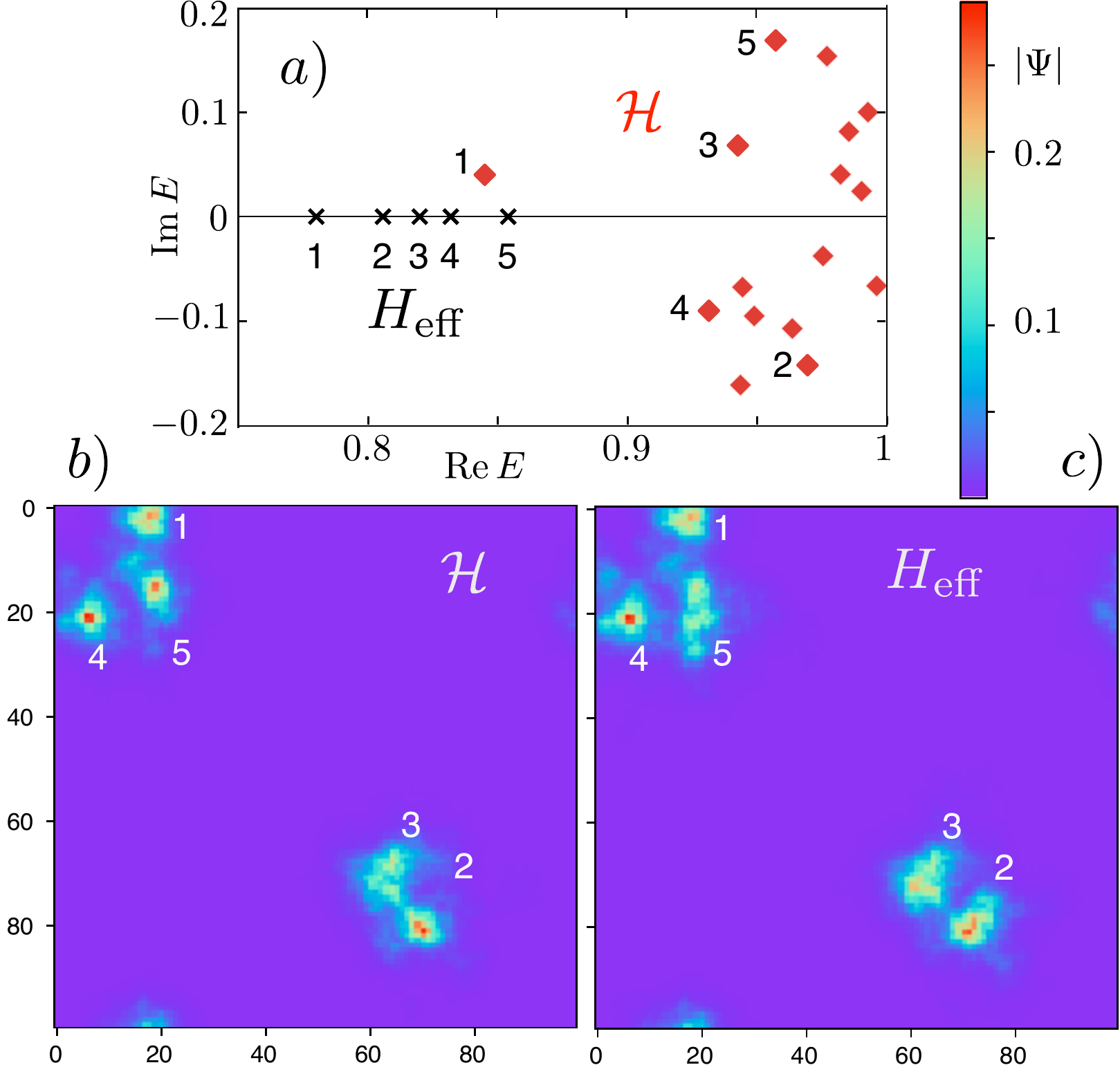}}
\caption{Energy levels (panel \textit{a}) and localized eigenstates (panels \textit{b,c}) of the non-Hermitian Hamiltonian ${\cal H}$ from Eq.\ \eqref{calHnonHermitian} and its Hermitian counterpart $H_{\rm eff}$ from Eq.\ \eqref{HeffHermitian}. The calculations are performed on a 2D square lattice (lattice constant $a\equiv 1$, band width $W_0=8$, periodic boundary conditions) for potentials $V_1$ and $V_2$ randomly and independently chosen at each site, uniformly in the interval $(-1,1)$. A constant offset $V_0=1$ was added to $V_1$ in order to ensure a positive $V_{\rm eff}$. The mapping from ${\cal H}$ to $H_{\rm eff}$ preserves the spatial location of the localized states, while the ordering of the energy levels $|E_n|$ in absolute value is changed. Panels \textit{b,c} show the eigenstates of the five lowest energy levels of $H_{\rm eff}$ and the corresponding eigenstates of ${\cal H}$. The locations are preserved but $E_2$ of ${\cal H}$ is pushed to higher absolute values.
}
\label{fig_nonHerm}
\end{figure}

The non-Hermitian Anderson Hamiltonian \cite{Tzo19,Hua19}
\begin{equation}
{\cal H}=-\nabla^2+V_1(\bm{r})+iV_2(\bm{r})\label{calHnonHermitian}
\end{equation}
has been studied in the context of a random laser \cite{Wie08}: a disordered optical lattice with randomly varying absorption and amplication rates, described by a complex dielectric function $V_1+iV_2$. On a $d$-dimensional square lattice (lattice constant $a$), the discretization of $-\nabla^2\mapsto a^{-2}\sum_{i=1}^d(2-2\cos k_i a)$ produces a spectral band width of $W_{0}=4d/a^2$. 

The Hermitian Hamiltonian
\begin{equation}
H_{\rm eff}=-\nabla^2 + V_{\rm eff},\;\;V_{\rm eff}=|\tfrac{1}{2}W_{0}+V_1+iV_2|-\tfrac{1}{2}W_0,\label{HeffHermitian}
\end{equation}
is positive-definite if $V_{\rm eff}(\bm{r})>0$ for all $\bm{r}$. The transformation from complex $V$ to real $V_{\rm eff}$ does not change the landscape function, because $\overbrack{\cal H}={\overbrack{H}}_{\rm eff}=H_{\rm eff}$. The localization landscapes are therefore the same and we would expect the eigenstates \cite{note3} of ${\cal H}$ and $H_{\rm eff}$ to appear at the same positions, provided that $V_{\rm eff}>0$. This works out, as shown in Fig.\ \ref{fig_nonHerm}.

\emph{Conclusion and outlook ---}
We have shown that the comparison matrix $\overbrack{H}$ provides a route to the landscape function for Hamiltonians that are not of the Schr\"{o}dinger form $H=-\nabla^2+V$. We have explored Hamiltonians for massive or massless Dirac fermions, with or without superconducting pairing. The broad generality of the approach is highlighted by the application to the non-Hermitian Anderson Hamiltonian.

The localization landscape can be used as a tool to quickly and efficiently find low-lying localized states in a disordered medium, since the landscape function $u(\bm{r})$ is obtained from a single differential equation $\overbrack{H}u=1$. These applications have been demonstrated for the Schr\"{o}dinger Hamiltonian \cite{Fil17,Pic17,Li17,Cha19}, and we anticipate similar applications for the Dirac Hamiltonian in the context of graphene or of topological insulators.

The comparison matrix offers a conceptual insight as well: Since equimodular Hamiltonians have the same comparison matrix, they form an equivalence class that localizes at the same spatial positions. This notion is distinct from the familiar notion of ``universality classes'' of Anderson localization \cite{Eve08}, which refers to ensemble-averaged properties. The equivalence class, instead, refers to sample-specific properties.

As an outlook to future research, it would be interesting to extend the approach from wave functions to energy levels. This has been recently demonstrated for the Schr\"{o}dinger Hamiltonian \cite{Arn19}, where the peak height of the localization function predicts the energy of the localized state. The correlation between peak heights and energy levels evident in Fig.\ \ref{fig_Rashba1D} suggests that the comparison matrix has this predictive power as well. Another direction to investigate is to see if the comparison matrix would make it possible to incorporate spin degrees of freedom in the \textit{many-body} localization landscape introduced recently \cite{Bal19}.

\emph{Acknowledgements ---}
The 2D numerical calculations were performed using the Kwant code \cite{kwant}. We have benefited from discussions with I. Adagideli and A. R. Akhmerov. This project has received funding from the Netherlands Organization for Scientific Research (NWO/OCW) and from the European Research Council (ERC) under the European Union's Horizon 2020 research and innovation programme.

\newpage

\appendix
\setcounter{page}{1} 

\section{Derivation of the comparison inequality}

The comparison inequality \eqref{Hcomparison} is derived by Ostrowski \cite{Ost37}. Here we give an alternative derivation, to make the paper self-contained.

In the most general case the matrix ${{H}}$ is a complex matrix, not necessarily Hermitian. We will initially assume that the diagonal elements ${{H}}_{nn}$ are real $\geq 0$ and relax that assumption at the end.

Decompose ${{H}}=\lambda\openone-L$, with $\lambda>\max_n {{H}}_{nn}$, so that the diagonal elements of $L$ are all positive. If we denote by $|L|$ the elementwise absolute value of the matrix $L$, one has
\begin{equation}
\lambda\openone-|L|=\overbrack{{H}},
\end{equation}
under the assumption that ${{H}}_{nn}\geq 0$.

Consider the Euclidean propagator $e^{-{{H}}t}$ for $t\geq 0$, and start from the inequality
\begin{equation}
\left| \textstyle{\sum_{m}}\left(e^{-{{H}}t}\right)_{nm}\Psi_m \right| \leq \textstyle{\sum_{m}}\left| \left(e^{-{{H}}t}\right)_{nm}\right| |\Psi_m|.\label{app_firstineq}
\end{equation}
We expand $e^{-{{H}}t}$ in a Taylor series,
\begin{align}
&\left| \left(e^{-{{H}}t}\right)_{nm}\right|=e^{-\lambda t}\left|\sum_{p=0}^\infty\frac{t^p}{p!}(L^p)_{nm}\right|\\
&\leq e^{-\lambda t}\sum_{p=0}^\infty \frac{t^p}{p!}(|L|^p)_{nm}=e^{-\lambda t}\bigl(e^{|L| t}\bigr)_{nm}=\bigl(e^{-\overbrack{\scriptstyle{H}}t}\bigr)_{nm}.\nonumber
\end{align} 
Substitution into Eq.\ \eqref{app_firstineq} gives
\begin{equation}
\left| \textstyle{\sum_{m}}\left(e^{-{{H}}t}\right)_{nm}\Psi_m \right|\leq\textstyle{\sum_{m}}\bigl(e^{-\overbrack{\scriptstyle{H}}t}\bigr)_{nm}|\Psi_{m}|.\label{app_secondinequality}
\end{equation}
This may also be written more compactly as
\begin{equation}
|e^{-{{H}}t}|\leq e^{-{\overbrack{\scriptstyle{H}}}t},\label{app_secondineq}
\end{equation}
with the understanding that the absolute value and inequality is taken elementwise.

If we now assume that all eigenvalues of $\overbrack{{H}}$ have a positive real part, then we may integrate both $e^{-{{H}}t}$ and $e^{-\overbrack{\scriptstyle{H}}t}$ over $t$ from 0 to $\infty$. On the one hand we have,
\begin{equation}
\int_0^\infty  e^{-{{H}}t}\,dt={{H}}^{-1},
\end{equation}
and on the other hand, in view of Eq.\ \eqref{app_secondineq}, we have
\begin{equation}
\left|\int_0^\infty  e^{-{{H}}t}\,dt\right|\leq\int_0^\infty|e^{-{{H}}t}|\,dt\leq \int_0^\infty e^{-\overbrack{\scriptstyle{H}}t}\,dt=\overbrack{{H}}\,^{-1}.
\end{equation}
We thus arrive at the desired comparison inequality \eqref{Hcomparison},
\begin{equation}
|{{H}}^{-1}|\leq\overbrack{{H}}\,^{-1}.\label{app_Hcomparison}
\end{equation}

The assumption that ${{H}}_{nn}$ is real $\geq 0$ can be removed my multiplying ${{H}}$ with the diagonal matrix
\begin{equation}
D_{nm}=\delta_{nm}\,e^{-i\,{\rm arg}\,{{H}}_{nn}}
\end{equation}
(setting $D_{nn}=1$ if ${{H}}_{nn}=0$). This matrix multiplication changes neither the comparison matrix, $\overbrack{D{{H}}}=\overbrack{{H}}$, nor the absolute value of the inverse, $|(D{{H}})^{-1}|=|{{H}}^{-1}D^{-1}|=|{{H}}^{-1}|$, hence Eq.\ \eqref{app_Hcomparison} still holds. Only the assumption of positive-definite $\overbrack{H}$ remains.


\begin{thebibliography}{99}
\bibitem{note4} The staggered discretization \eqref{eq_staggering} corresponds to the tight-binding Hamiltonian $H=(\hbar v_{\rm F}/a)\sigma_x\sin ka+(\hbar v_{\rm F}/a)(1-\cos ka)\sigma_y+V\sigma_0+\mu\sigma_z$, which gives the dispersion relation \eqref{dispersionrelation} when $V=\mu=0$.
\bibitem{note3} Because ${\cal H}^\dagger={\cal H}^\ast$, the left and right eigenvectors are each others complex conjugate and we do not need to distinguish between these when plotting the absolute value in Fig.\ \ref{fig_nonHerm}b.
\end{thebibliography}
\end{document}